# Disrupted Routines Anticipate Musical Exploration

**Authors:** Khwan Kim[1], Noah Askin[2], James A. Evans[3,4]

**Affiliations:**

[1] INSEAD, Boulevard de Constance, 77300 Fontainebleau, France.

[2] The Paul Merage School of Business, University of California–Irvine, 4291 Pereira Drive, Irvine CA 92697.

[3] Department of Sociology, University of Chicago, 1126 E 59th St, Chicago, IL 60637.

[4] Knowledge Lab, University of Chicago, 5735 South Ellis Avenue, Chicago, IL 60637.

**Understanding and predicting the emergence and evolution of cultural tastes manifest in consumption patterns is of central interest for social scientists, analysts of culture[1–3] and purveyors of content. Prior research suggests that taste preferences relate to personality traits[4–6], values[7], shifts in mood[8–10], and immigration destination[11–13], but understanding everyday patterns of listening and the function music plays in life have remained elusive, despite speculations that musical nostalgia may compensate for local disruption.[14] Using more than a hundred million streams of 4 million songs by tens of thousands of international listeners from a global music service catering to local tastes, here we show that breaches in personal routine are systematically associated with personal musical exploration. As people visited new cities and countries, their preferences diversified, converging towards their destinations. As people experienced COVID-19 lock-downs, and then again when they experienced reopenings, their preferences diversified further. Personal explorations did not tend to veer toward the global average, but away from it and toward distinctive regional musical content. In all of these settings, musical preference reflected rather than compensated for life's surprises. We explore the relationship between these findings and global patterns of behavior and cultural consumption.**

**Main text**

Understanding and anticipating the emergence and evolution of cultural tastes and consumption patterns remains the holy grail for analysts of culture[1–3] and companies that deliver cultural content: Netflix has held contests to identify who could design a better recommendation algorithm for suggesting content to its users, TikTok has raced to the top of the mobile application world with "addictive" short video recommendations that better capture and align with users' tastes, and digital streaming platforms (DSPs) in music compete on both quality of recommendation engine and size of available library. Artists, companies, and scholars alike seek to understand and influence users' tastes—an endeavor made easier by the increase in digital trace data these services generate[15–18]. Yet examining what shapes taste preferences remains a challenge, as they change over time and across contexts, and as the quantity of available culture explodes. Aside from the rise in mood- and context-based playlists for music (e.g., "Mood Booster", "Rainy Day" or "Morning Commute" playlists on Spotify), a trend that aligns with evidence showing that musical preferences are associated with context and mood[8,9], there is little evidence demonstrating how tastes are influenced by everyday shifts in routine and context. We propose and demonstrate that disruptions in routine and shifts in context vary positively with consumption and listening habits.

Here we build on work linking cultural taste preferences to consumers' intrinsic characteristics[4–7] and external environment[11–13] by characterizing both travel–vacations, business trips, etc.–and Covid-19-related lockdowns (and re-openings) as disruptions in consumers' routines that influence consumption behaviors. We contribute to work showing how tastes evolve[19] as well as the functions played by music[20–22], particularly in new settings. Concretely, we investigate monthly patterns of spatial routine change, as measured by geographical distances traveled by consumers, and their impact on consumers' tendency to explore outside of their usual cultural consumption patterns, measured as the distance between their past taste and their new monthly listening behavior. We also explore the lockdowns and reopenings associated with Covid-19 as potential drivers of changes in consumption patterns.

**The construction of taste vectors for preference measurement**

To systematically evaluate the impact of geographic and routine changes on listening habits, we obtained extensive listener data from Deezer, the 7$^{th}$ largest music streaming service globally (18 million users, including 10 million paying users) whose strategy has focused on identifying and streaming regional music in addition to global hits. The raw data from Deezer consist of complete listening histories from January 2018 through December 2020 for 44,794 randomly selected users across nine countries—France, Germany, the UK, Brazil, Australia, Russia, South Africa, Morocco, and Mexico. Our sample includes approximately 10,000 users from each of the first four countries, whose subscriber bases are relatively large, and about 1,000 users from each of the other five countries with smaller subscriber accounts. Female users account for 27.3% of the total. The data include over 596 million streams of 4,276,197 unique songs. **Fig. 1a** shows our focal countries, the number of users sampled in each country, and the number of streams from those users during our observation period. **Fig. 1b** provides an excerpt of the data collected for each streamed song.

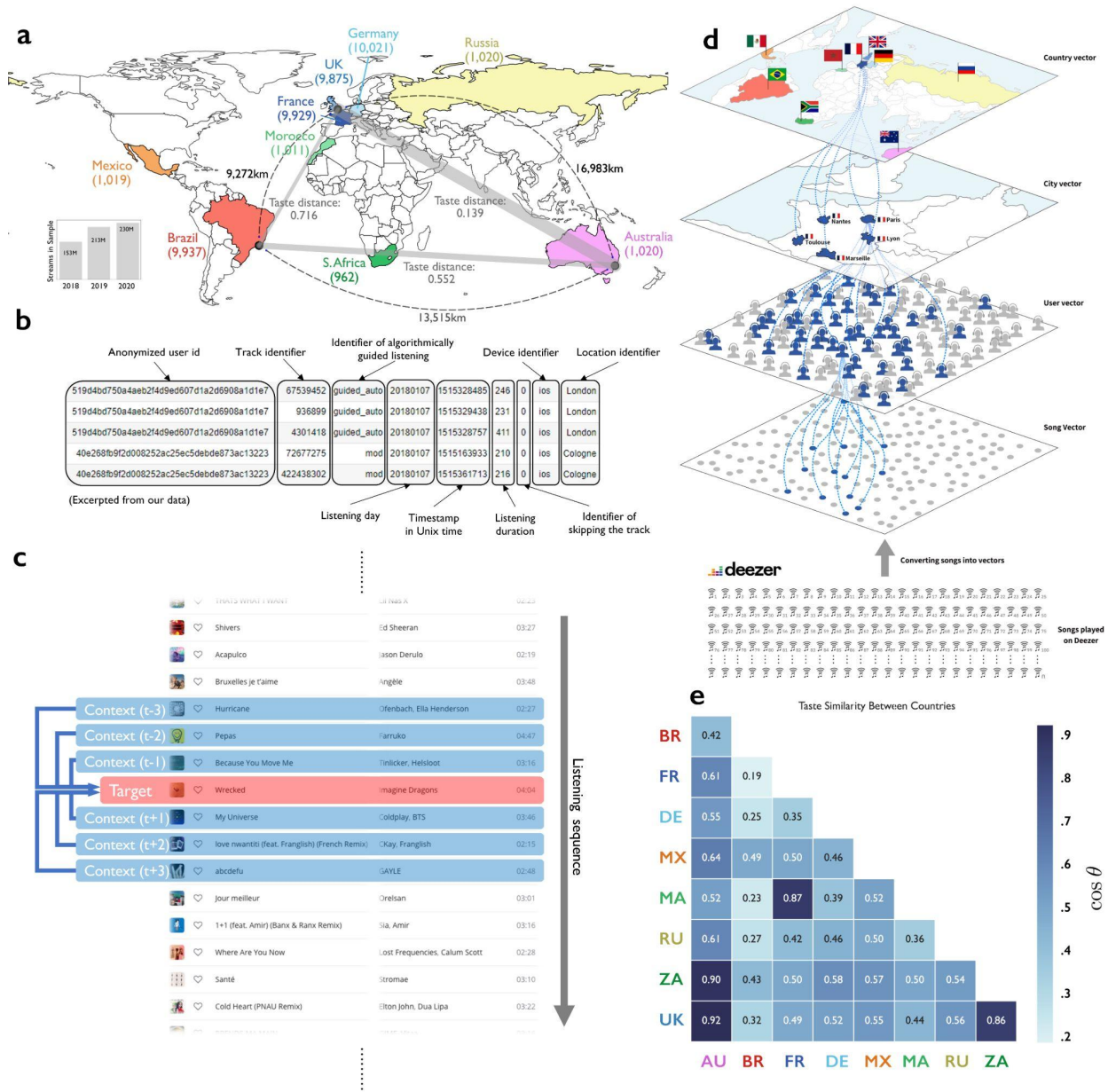

**Fig. 1. Data and measurement overview. a,** Countries and continents of users in sample, number of musical streams per year, and distinctions between geo-political distance, and musical/taste distances emergent from streaming songs-in-context. **b,** Annotated sample of streaming data fields. **c,** Arrangement of streamed songs in order to determine context for Song2Vecor (S2V) embedding. **d,** The construction of S2V-based distances between songs, people, which represent the centroid of listened song vectors, cities, which represent the centroid of listening people there, and countries, which represent the centroid of listening within cities, around the world. **e,** S2V cosine similarity between country-level taste vectors: Australia (AU) and the United Kingdom (UK) are closest (.92), while France (FR) and Brazil (BR) are furthest (.19).

Detailed streaming data allow us to create taste profiles that represent individuals, cities, and countries over time. We created these profiles using a neural network model previously developed to capture song-embeddings in the form of a numeric vector that represents a song's relative position in a "playlist", or the sequence of songs to which a user listens. Inspired by collaborative filtering algorithms that exploit songs' play sequence information[23], we apply the Word2Vec algorithm, designed to identify word meaning from context,[24,25] in order to identify song similarity from context. Specifically, we draw on users' listening sessions containing multiple songs to produce embeddings for songs in a latent space. We call this song-embedding Song2Vec (henceforth S2V). The overall process of constructing S2V is summarized in Methods.

With S2V, we measure qualitative similarity or dissimilarity between songs–similar to the way Word2Vec enables measurement of semantic distance between word vectors. As embeddings can be aggregated at different levels of analysis (e.g., user, city, country), S2V is also capable of capturing cross-level relations (e.g., song-user, song-city, user-city, user-country, etc.) and within-level relations (e.g., user-user, city-city, and country-country relations). We define each user's taste vector as the multivariate mean or centroid of S2V vectors consumed by a given user. In a similar manner, we define a city's taste vector as the mean of user taste vectors for users living in that city. Finally, we define a country's taste vector as the mean of city taste vectors for all cities located in that country. **Fig. 1c** illustrates the process we use to create them.

In our analysis, we distinguish taste distance from geographical distance. We measure geographical distance between any two cities or countries using the haversine formula, which determines great-circle distances given their longitudes and latitudes, whereas we measure taste distance via the cosine distance between two playlist vectors. **Fig. 1a** shows the geographical and taste distances between London, Rio de Janeiro, and Sydney. Although London is physically closer to Rio de Janeiro than to Sydney, it is much closer in taste to Sydney. Similarly, we can calculate taste similarity between geographies by subtracting the cosine distance between two locations from 1. **Fig. 1e** displays taste similarities between countries.

Our primary interest is in assessing the effect of spatial and routine disruptions on the evolution of individuals' cultural tastes. For the geospatial movement of each listener, we first capture their home city–the city in which they listened to the most music in a given year. Then we measure the physical distance between that home city and all other cities they visit in a given month. This distance is at its lowest (i.e., zero) when a listener stays in their home city for the month, while it is much higher when they make international trips. For the cultural taste change of each listener, we measure how much a listener consumes songs qualitatively different from those they listened to in past, which we capture by calculating the cosine distance between the current month's user vector and their vector from the prior six months. Taste distance is lowest (i.e., close to zero) when a user listens to a set of songs on repeat that they regularly listened to over the prior half-year, but is much higher (i.e., close to one) if they explore entirely new songs from genres and artists that dramatically different from their usual listening habits.

**Travel distance and taste exploration**

Regression analysis of users' taste evolution at both global and country levels shows that the greater the user's geospatial movement, the greater the disruptive evolution of their taste profile.

We consistently observe that over the three years from 2018 to 2020, those who experienced greater change geospatially were more likely to consume cultural products novel to them (bold navy line in **Fig. 2a** and Model 3-7 in **Extended Data Table 1**). A similar pattern appears across countries, although the effect in some small-sample countries is not statistically significant. Nevertheless, we see a significantly positive association between geospatial change and taste exploration in France, UK, Germany, Brazil, Russia, and South Africa (inset plot in **Fig. 2c** and Model 1-9 in **Supplementary Table 2**). Routine disruption–here in the form of travel–appears to prompt users to try something new.

We further examine the interaction effect between routine disruption and *taste deviation from global taste* to evaluate whether the positive effect of geospatial movement on taste exploration is more salient when a user deviates from popular, mainstream taste. Our regression analysis supports our prediction: the likelihood of taste exploration as a function of geospatial disruption increases with individuals' deviation from global taste (see Model 8-11 in **Extended Data Table 1**). We visualize this interaction effect in **Fig. 2b**.

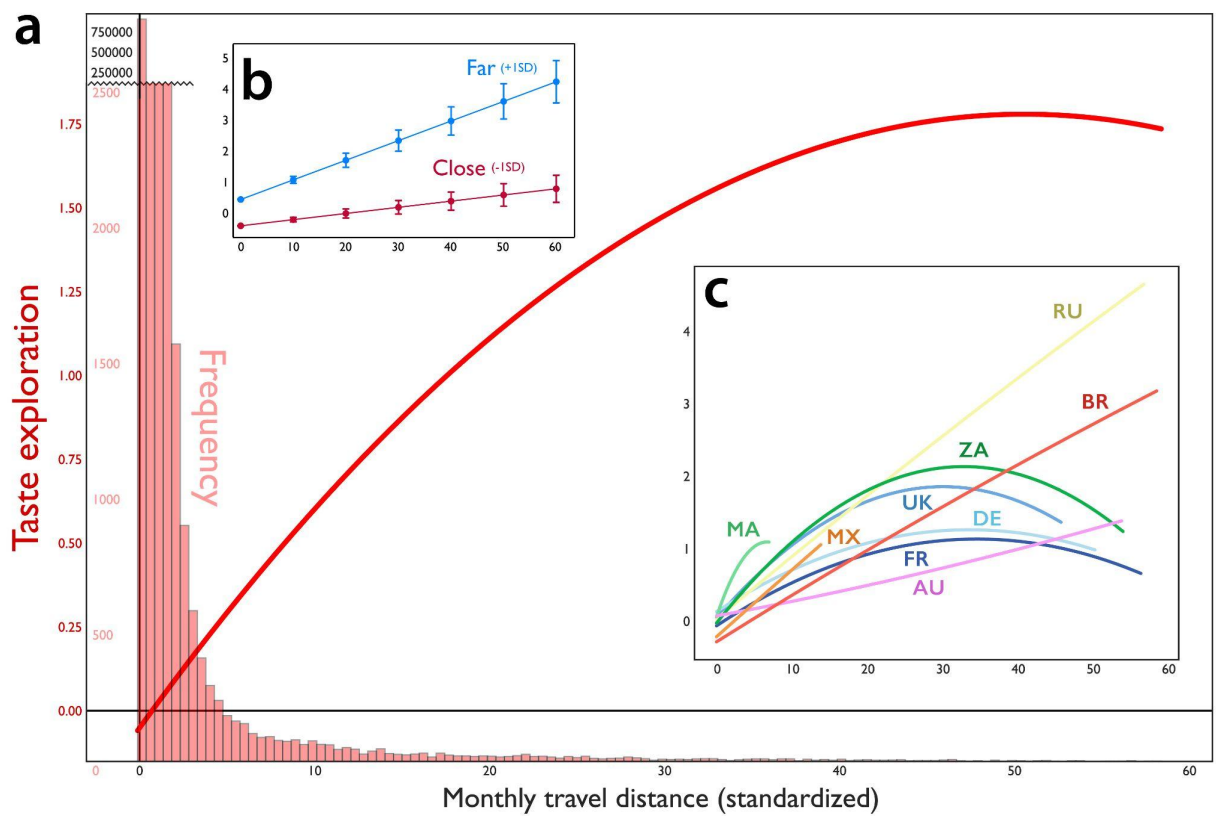

**Fig. 2. Travel distance and taste exploration. a**, Physical distance from listeners' "home city" has a positive association with their monthly taste exploration, suggesting that the diversity of their listening preferences scales with their geographical exploration. **b**, This effect positively interacts with their tendency to deviate from global taste. A listener is measured as distant from global taste if her taste vector is one standard deviation beyond the average distance from the global vector in a given month as measured by cosine distance, and close if one standard deviation closer. This analysis demonstrates that the positive effect of travel distance on taste exploration is higher among deviants than conformists. **c**, Same as Fig. 2a, but for each country.

**Taste distance and adaptation**

Related to the relationship between geospatial change and taste exploration is the possibility that people tend to assimilate their cultural taste towards the dominant taste in their new/foreign environment, a tendency likely amplified with greater distance between hometown taste and that of the new environment to which they are exposed. We examine this relationship by comparing how much a user's listening habits move towards the prevailing listening trends of another city as a function of how far that city's taste is from the user's home city (see Methods for details).

We find that the greater the inter-city taste distance, the stronger the tendency of a user's listening to adapt to the new environment (**Fig. 3a, b** and **Extended Data Table 2**). This is illustrated in an example of a Londoner who travels to Rio de Janeiro and Johannesburg in **Fig. 3d**. Both cities have comparable geospatial distance from London, while their taste distance from London varies considerably. Our results suggest that this Londoner is more likely to adjust her taste to Rio de Janeiro than Johannesburg because the difference in the tastes between cities is greater when traveling to Rio de Janeiro than Johannesburg.

We further examine the interaction effect between taste distance and geographical distance to evaluate whether the positive effect of taste distance on taste adaptation is more salient when a user spatially deviates further from home. Our regression analysis supports our expectation that the linkage between taste adaptation and taste distance increases with listeners' geospatial distance from their home cities (see Model 8-11 in **Extended Data Table 2**). We visualize this interaction effect in **Fig. 3c**.

**Routine disruption and taste exploration**

The global pandemic that began in early 2020 provides us with a unique opportunity to examine the impact of a different kind of routine disruption on consumption patterns. Governments took a wide range of measures in response to the COVID-19 outbreak as it flared up or simmered down. Throughout 2020, most of the countries in our data imposed strict nationwide confinement and lifted it several weeks later, although they varied in the stringency and timing of their intervention. We make use of the fact that the two crucial policy responses—lockdown and reopening—disrupted individuals' geospatial routines. Our interest here is in whether lockdown and re-opening, which are themselves varieties of geographical routine change, were similarly associated with changes in taste exploration.

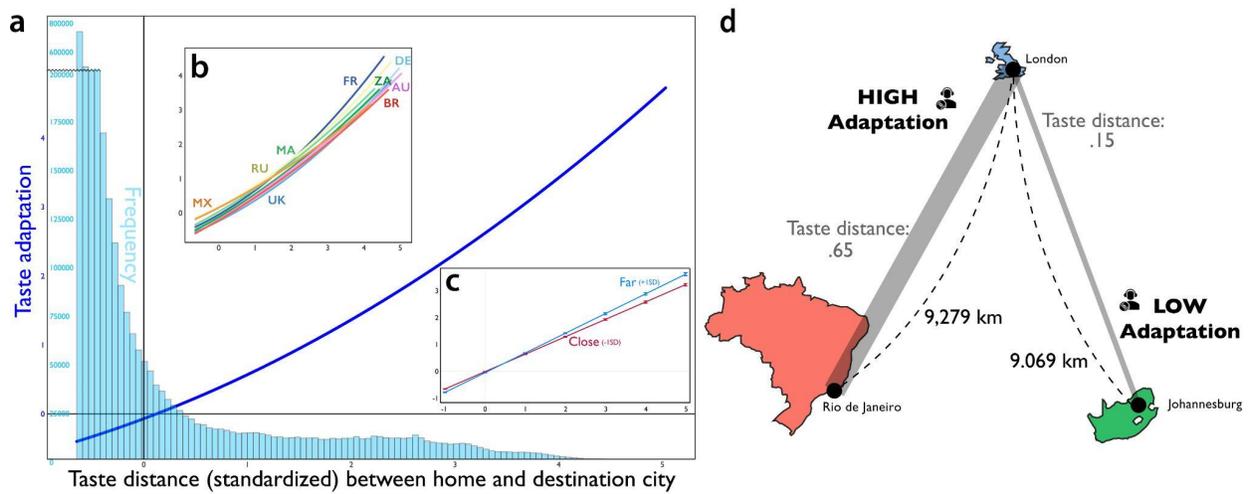

**Fig. 3. Inter-city taste distance and user taste adaptation. a,** A positive association between taste distance from listeners' "home city" and their destination city suggests that their adaptation to local listening preferences scales with cities' taste distance, globally, and **b**, for each country in our sample. **c**, This effect of taste distance is positively moderated by geospatial distance, where "far" apart cities are one standard deviation greater than average inter-city distances, and "close" cities are one standard deviation smaller than average inter-city distances. **d,** Example of the average effect in the context of listeners traveling between London and Rio de Janeiro versus Johannesburg.

Users' taste exploration and travel are relatively consistent from July 2018 through December 2019 (see **Fig. 4a**) with taste divergences around the winter holidays, reflecting increased consumption of holiday music typically not listened to the rest of the year, though trends vary by country (see also **Extended Data Fig. 1**). Travel also manifests modest increases around winter and summer holidays. Exploration and travel decouple as the pandemic breaks out in early 2020. Taste exploration spikes during March and April—months during which the first and strictest lockdowns were implemented—reflecting shifts in people's tastes when day-to-day routines are disrupted. Taste exploration spikes again directly following the time many countries reopened in May-June, 2020.

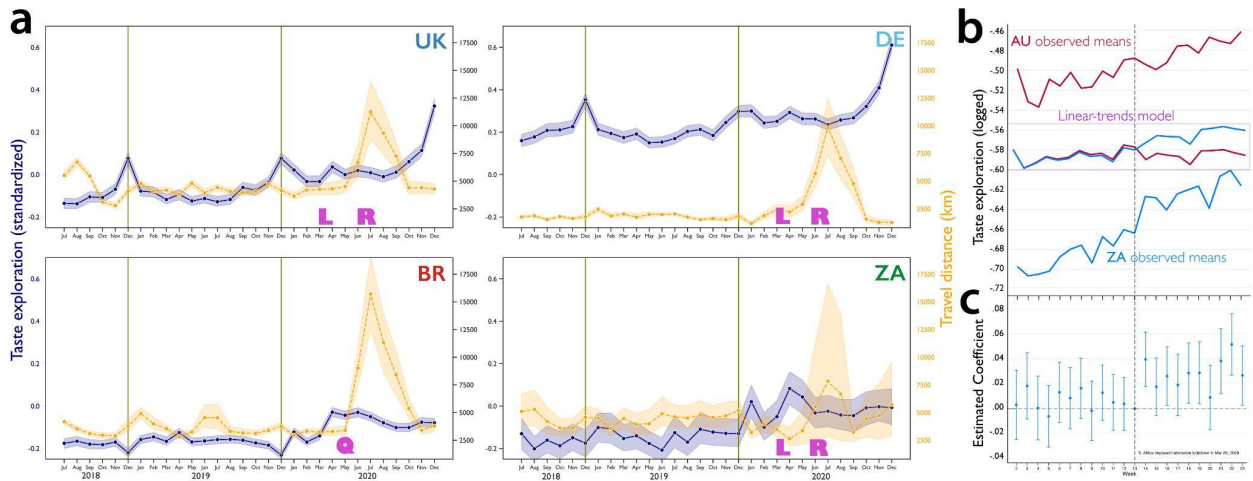

**Fig. 4. Taste exploration and travel distance. a**, Monthly taste exploration and travel distance surrounding Covid-19 lockdown (L), reopening (R), and regional quarantine (Q) events for the United Kingdom (UK), Germany (DE), Brazil (BR), and South Africa (ZA). **b**, Parallel trends prior to South Africa's nationwide lockdown in taste exploration between South Africa and Australia. While the observed means of Australia and South Africa show parallel trends prior to the South African lockdown, the latter starts to increase post-lockdown. The linear-trends model further confirms the parallel-trends assumption. **c**, Week-specific treatment effects of lockdown on taste exploration. This visualizes the result of the Granger causality test that augments our Differences in Differences (DiD) model to include dummies as if treatment had occurred in the past. It fits a generalization of the DiD model and plots the estimated coefficients with 95% CIs. The contrast between pre- and post-treatment periods in coefficients is aligned with the result of the Granger causality test ($F=0.86$, $p$-value=0.583) obtained by performing a joint Wald test on the coefficients.

This pattern—a lockdown-related drop in travel followed by reopening, each coinciding with increased taste exploration—suggests a connection between spatial routine disruption and the evolving diversification of cultural taste (**Fig. 4a** and **Extended Data Fig. 1**). We examine this connection in greater detail via difference-in-difference (DiD) statistical analysis (**Fig. 4b,c**; **Extended Data Table 3**).

DiD analyses yield separate estimates for selection and treatment effects, enabling us to examine the effects of treatments like lockdowns and reopenings over time across treated and comparison groups. We highlight South Africa and Australia, similarly represented in our data and which share a strong taste similarity (see **Fig. 1e** and **Extended Data Fig. 2**) In response to the initial outbreak of COVID-19 in 2020, South Africa introduced strict, swift measures that disrupted routines much more dramatically than Australia did[26]. The Stringency Index (SI) calculated by Oxford COVID-19 Government Response Tracker[27] captured the difference between the two countries. Across the month of March 2020, South Africa's SI increased from the second to the 89th percentile, while Australia's moved only from the 11th to the 31st percentile. South Africa imposed a 21-day national lockdown with the deployment of military force, effective from 26 March, but Australia never went into complete lockdown. The federal government focused on international travel restrictions[26], leaving implementation of major measures to the discretion of state governments. **Extended Data Fig. 3** visualizes the SI of the two countries during the first pandemic period.

For our DiD estimation, we calculate differences in taste exploration before and after the imposition of a nationwide lockdown for both countries, visualized in **Fig. 4b**, then calculate the difference between these differences across the two countries (see **Methods**). Our main model (see **Fig. 3c** and our Model 7 in **Extended Data Table 3**) reveals a strong, positive average treatment effect on the treated group (ATET) ($\beta$=0.023, $p$-value<0.001, two-tailed), suggesting that South African users who underwent drastic routine disruption due to the national lockdown significantly increased their taste exploration during that lockdown compared to the Australian users whose routine disruption was less severe.

**Discussion**

Across settings and contexts, musical preferences reflected rather than compensated for life's surprises. Disrupted routines as a function of local and international travel were systematically associated with increases in exploration and cultural curiosity, influenced by the locations listeners traveled. Personal explorations did not veer toward the global center of taste, but away from it and toward distinctive regional musical content. Furthermore, we find that disruptions associated with COVID-19 related lock-downs, and then reopenings, both appear associated with greater musical exploration. Collectively, listeners appear to construct a soundtrack that mirrors the disruptions of their lives rather than one that counterbalances them.

Our study and findings have several limitations. Some of the countries in our sample had uneven listener samples, and there were limits to the amount of global diversity in response to the pandemic we could use as the basis for natural experiments. The patterns raise causal identification questions, but in some cases involve forced actions (e.g., lockdowns) imposed exogenously, which are much more likely to influence listening preferences than the converse. Nevertheless, these findings link shifts in convention with changing preferences in cultural consumption. They pose novel measurement opportunities that arise from consistent signals that cultural consumption reveals about the lived experiences and inner states of those who consume them.

**Methods**

**Deezer, The Listener Dataset, and Song-Embeddings (S2V)**

Here, we provide details of the Deezer listener dataset and the analyses conducted. The increasing use of music streaming services and recent availability of data on music consumption on streaming platforms have enabled in-depth research on cultural consumption at global[28], within-country[29], and cross-country levels[30]. To the best of our knowledge, however, no previous study has analyzed a large sample of real-time data on the listening behavior of global users over multiple years.[1]

Compared to Spotify, a top market player in music streaming, whose number of paying subscribers is 180 million, Deezer's user base is smaller, though it is available in more countries than Spotify (187 to Spotify's 184).[2] There is no significant difference between Deezer and Spotify with respect to the amount of music content available on the platform. Both platforms have massive music libraries with over 70 million tracks. Nevertheless, Deezer is distinctive in several ways. First, Deezer focuses on distinguishing its platform with localization, providing a large selection of exclusive, original foreign-language music and podcasts as opposed to exclusively global-reach, international content. For example, Deezer offers 32,000 local and international radio stations, while Spotify does not offer traditional radio on the app. Deezer is also more likely to meet music connoisseurs' high expectations for audio quality. Unlike Spotify, Deezer uses FLAC (Free Lossless Audio Codec) for paying users of the HiFi option, which allows those users to listen to music with CD-like audio quality. Furthermore, Deezer has a distinctive feature that facilitates music discovery. *SongCatcher* in Deezer identifies any song playing in the physical vicinity of a listener. It helps users quickly catch and save music playing nearby without leaving the app. This feature provides detailed information about the song followed by the option to add it directly to the user's library. In sum, with these unique features, Deezer may appeal more to music consumers who value localized, high-quality audio, and the discovery of new sounds.

---

[1] Real-time data in our setting refer to user's listening log that is delivered immediately from user's device to the server of the app.
[2] As of January 2022.

Of the distinctive characteristics in Deezer, the greater localization of content can be seen, in part, via cross-national diversity in the top music charts across different countries on the app (i.e., the degree to which local songs appear on the chart for most listened music, relative to global hit songs). Cross-national diversity or localization is low if many of the same songs appear across different countries' Top 100 charts, while it is high when more locally unique songs appear on those charts. We collected the daily most popular 100 songs from Deezer and Spotify in the nine countries we sampled for seven days from December 12 to 18 in 2021. Using the Jaccard similarity coefficient that captures common elements shared by two groups, we show how much the same songs appear across different countries' Top 100 charts on Deezer and compare the cross-country similarity with Spotify's cross-country similarity (see **Extended Data Fig. 2**).

Among country similarities, the lowest Jaccard coefficient is found between France and Brazil, meaning that the French listeners' music tastes collectively differ from Brazilian tastes more than any other pairing among the nine sampled countries. The three English-speaking countries—the United Kingdom, Australia, and South Africa—share high similarities, likely associated with the linguistic influence that allows for the easy diffusion of foreign music and the emergence of similar taste clusters. This pattern is also seen in the high similarity between Moroccan and

French taste–approximately one-third of Moroccans speak French, and our data reveals that many Moroccan listeners tend to stay in France either for a short (a few months) or long (more than a year) term over the course of our sample. Furthermore, the lower average Jaccard coefficient in Deezer compared to Spotify means that Deezer listeners are more likely to consume local songs over global hits than Spotify users.

In terms of listener demographics, of the 44,794 users in our data, the largest age group comprises those in their 20s (10,835 users), accounting for 24.2% of the total sampled users. The age-gender distribution of the sampled users in each country is illustrated in **Extended Data Fig. 4**, with 23.7% women listeners.

A variety of information on users' listening is automatically sent from the listening device to Deezer's internal server and stored in a log file. Our primary data is derived from those detailed logs of listening behavior at the individual user level. The listening log data has nine components: (1) anonymized user identifier, (2) track identifier, (3) origin of the listen (e.g., whether playing a song from your collection, through an editorial playlist, or in-app recommendation algorithm), (4) listening date, (5) listening time stamp, (6) duration of the listen, (7) skip identifier (i.e., whether you click next before the track ends), (8) platform used (e.g., Android, iOS, web, etc.), and (9) geographical location of the listener at the city level. We use all these elements except platform information (8) to create our central variables.

**Construction of S2V Embedding**

The data consists of 63.3 million listening sessions of 496 million songs streamed by our users after dropping short-listened streams (i.e., shorter than one minute). In defining a listening session, we assume that a current listening session, in which a user consecutively listens to multiple songs in a row, ends if there is a break longer than 5 minutes between the time when the last stream ends and the time when the next stream starts. Session demarcation is important because each session may entail a unique context. For example, a user who jogs along the river in the morning listening to uptempo dance music may play downtempo folk music on her way to work after a considerable session break, such as showering or eating breakfast. As such, our 5-minute threshold takes into consideration the possibility that a user's listening context may change even after a short pause with respect to mood, emotion, situation, etc.

Once we identified the listening sessions, we created our S2V model. Hyper-parameters of the model are set as follows in the Gensim implementation of W2V: dimension (i.e., vector size = 300, minimum count of songs = 2, iteration = 100, skip gram = FALSE, negative sample size = 20, context window = 3). In other words, the dimensionality of the feature vectors is 300. The maximum distance between the "target" song and its neighboring "context" songs is set to 3, meaning that we train our model such that the vector of the target song is influenced by its three preceding and three following songs. Songs must appear at least twice in our data to be included in the training sample. The CBOW (Continuous Bag of Words) model is used instead of the Skip-gram model as the training algorithm because the even though the number of songs and streams is large in total, the diversity of songs played beside one another means we need an algorithm robust to data sparsity[31]. In addition, 20 negative sample songs (i.e., "noise songs" never played within six songs in a playlist together) are drawn to optimize the quality of the resulting vectors. Even though negative sampling induces statistical bias making it inappropriate for inference, it works well in practice for prediction in word embeddings[32], question answering[33], and many other applications, outperforming the comparable but unbiased contrastive learning approach by co-minimizing bias and variance[32]. The result of this operation, run for all streamed songs across our entire sample is a 300 dimensional manifold that embeds songs as a function of their proximity within context-specific playlists. We then assign a unique 300-dimensional vector to each song as a function of their position relative to other sings from our data within the global space of contextual listens.

**S2V Embedding Validation**

We initially validated the S2V measure by sampling a focal song and searching for its most similar songs based on the cosine similarity (i.e., its S2V). Then, we evaluated whether the focal song was more analogous to similar songs musically and contextually than randomly selected songs.[3] We repeated this procedure multiple times. We found that song vectors with higher cosine similarity values are more likely to be categorized as the same style or genre as the focal song, verifying the robustness of our song-embeddings (see **Extended Data Fig. 5**).

To illustrate our S2V's validity, we visualize four examples, each of which shows a focal song, its ten most similar and dissimilar songs as defined by our S2V model (see **Extended Data Fig. 5a-d**). Using a t-distributed Stochastic Neighbor Embedding (t-SNE), a nonlinear dimensionality reduction algorithm[34], we first reduce a 300-dimensional vector space to two dimensions. We picked one of the most popular Pop/Hip-Hop songs in 2018 (*God's Plan* by Drake), a long-running dance song released in 1983 (*Billie Jean* by Michael Jackson), a non-English Pop song by a Brazilian musician (*Bum Bum Tam Tam* by MC Fioti), and an chart-topping Alternative Rock song (*Creep* by Radiohead). Focal songs are marked in black, the most similar in blue, and the most dissimilar in red. Across the four examples, a focal song's most similar songs are predominantly by the same musician or by other contemporaneous musicians in the same genre, indicating that our S2V model captures both acoustic similarity and other time-varying aspects of songs. In addition, using the entire sample, we compare the level of average cosine similarity within artists and the level of average cosine similarity across artists, expecting the former to be higher than the latter (see **Extended Data Fig. 5e**). The former—within-artist similarity—represents how similar a focal artist's songs are to each other within that artist. The latter—across-artist similarity—reflects how similar a focal artist's songs are to all the other artists' songs in the population. The average within-artist similarity is 0.715, while the average cross-artist similarity is 0.491. The paired t-test further confirms the statistical significance of the difference between the two similarity scores (t-statistic=285.033; *p*-value=0.000).

---

[3] We compared the focal song and the other songs with respect to their acoustic (e.g., tempo, key, mode, etc.) and categorical (e.g., genre, release period, musician gender, etc.) features.

**Analysis of Taste Exploration (Fig. 2a-c; Extended Table 1; Supplementary Table 2 & 3)**

**Dependent Variables**

*Monthly taste exploration*. Our first dependent variable, *taste exploration*, captures the extent to which a cultural consumer explores novel tastes by measuring the distance between a user's taste vector in the current month and the mean of that user's taste vectors over the prior six months (i.e., up to the end of the last month). More formally, user *i*'s taste exploration in month *t*, $Taste\ Exploration_{i,t}$, is measured as the following:

$$\text{Taste Exploration}_{i,t} = \cos \text{dist} \left( UV_{i,t}, \frac{\sum_{t-6}^{t-1} UV_i}{6} \right)$$

where $UV_{i,t}$ is the user vector of listener $i$ in month $t$.

**Independent Variable**

*Monthly geospatial disruption (travel distance)*. Our key independent variable, *geospatial disruption*, captures the geographic distance a listener covers in a given month upon leaving her home city, assuming she used Deezer at least once during her trip. This first requires specifying one's home city. We infer a home city for each listener by detecting the city where they used Deezer most. We then identify which cities a listener traveled to while away from home. We measure the monthly travel distance by calculating the haversine distance between a user's home city and every city traveled to by that listener in a given month. Then, we sum those haversine distances per month, resulting in a listener's monthly total travel distance. For all analyses in which travel distance is interpreted or visualized, we standardize it into its z-score.

**Control Variables**

*Monthly algorithmic guided-ness*. Many music streaming services invest heavily in developing algorithms to provide song recommendations (e.g., "Recommended Playlist", "Made for Jane Doe", "Based on your recent listening", etc.) to their users. Typically, those algorithms generate a list of recommended music that satisfies a given user's taste but has not been discovered by the user. Nevertheless, it remains up to the user to play that playlist. Idiosyncratic behavioral tendencies to select and endure a playlist are likely to influence the extent to which one explores novelty. We, therefore, control for the degree to which a given user plays music based on algorithmic recommendations per month. It is measured as the ratio of the number of a given user's streams that came from algorithmic recommendations to her total streams per month.

***Monthly listening count***. We also consider the total number of a user's stream counts in a given month to control for the potential effect of listening frequency on a user's tendency to consume novel music.

***Monthly average song recency***. Some users have temporal preferences in their musical tastes. For example, a preference for newly released music may affect their propensity for taste exploration. We, therefore, control for the average song recency across a user's monthly listening sessions. We first calculate a listen-age by counting the number of weeks between a song's release and when a user listens to it. Then, we subtract the listen-age from the maximum song age in our data to create an inverse song age, which we define as song recency. Monthly average song recency is the mean of song recency for all streams by the user in a given month.

***Distance from global taste***. Taste exploration also likely varies with a user's tendency to converge toward or diverge from global trends such as the most popular songs in the world for a given month. We address this likelihood by controlling for each user's taste distance from the global taste vector—the cosine distance between the focal user's taste vector and the mean taste vector of all other users in our sample.

***Month dummies***. To control for time-varying trends, we include dummy variables for listening months in our model.

The longitudinal trends of the control variables except month dummies are visualized in **Extended Data Fig. 6**.

**Empirical Strategy and Results**

For the analysis of taste exploration, simple OLS regression is not appropriate for two reasons. First, the presence of repeated observations for the same users over time violates the assumption of observation independence. Second, the variance of the error terms might be heterogeneous across different cross-sectional units, yielding unmodeled heteroscedasticity. Therefore, we run a fixed-effects linear model using the *xtreg* command in Stata 17. We also use robust standard errors, clustering on the user to further account for the presence of repeated observations.[35,36] We standardize all of our variables into z-scores before running the model in order to make comparisons between their coefficients interpretable. The descriptive statistics of the variables are reported in **Supplementary Table 1**.

The global coefficient of monthly travel distance with respect to monthly taste exploration is visualized in **Fig. 2a,** whose inset figure shows the local coefficient for each country. Our focal coefficient is the linear term of travel distance, and it remains significantly positive across the models as shown in **Extended Data Table 1**. Its quadratic term is negative, suggesting a concave relationship of diminishing increase between travel distance and taste exploration: Given the differences in magnitude between the two terms, we interpret the results as indicating a positive association between geospatial movement and taste exploration.

We also find a positive interaction between travel distance and user's deviation from global taste. Its effect remains significant even if any single or pair of countries are removed from the data

(e.g., Brazil or/and Australia)[4] (see Model 8-11 in **Extended Data Table 1**). This suggests that the association of geospatial movement to taste exploration is positively moderated by a listener's tendency to move away from the global taste vector. In other words, one's exploration of novel taste when traveling is amplified in users with more nonconforming listening patterns. We report the results of country-level regression in **Supplementary Table 2 and 3**.

We test the robustness of our results against several different operationalizations of both the dependent and independent variables. For the dependent variable, *taste exploration*, we first run the same set of analyses after log-transforming it to assess whether our results are driven by any skewness of its distribution. Results remain substantively the same. Moreover, we apply different time windows for defining a listener's baseline taste to see if our results are robust to varied definitions of one's baseline taste structure.[5] Concretely, we construct listeners' baseline taste vectors with streams from only the month prior to a focal month, and we measure taste exploration by the cosine distance between last month's taste, $UV_{i,t-1}$, and the current month's taste, $UV_{i,t}$. We also consider the cumulative nature of taste formation by computing a cumulative average of a user's complete collection of past streams, since the beginning of our observation window (January 2018), up until *t*-1 month. We then measure their taste exploration by the cosine distance between that cumulative past taste, $\frac{\sum_{1}^{t-1} UV_i}{t-1}$, and the current month's taste, $UV_{i,t}$. In both cases, our main results remain substantively the same. For the independent variable, *travel distance*, we substitute the current operationalization—summing all travel distances by a listener in a given month—for the average of those travel distances. This does not substantively change our results.

---

[4] For the reported analysis, we drop Brazil because its national taste vector appears relatively far from the other eight countries (see **Fig. 1e**). We drop Australia because their travel pattern seems idiosyncratic compared to the other eight countries (see **Extended Data Fig. 1**). Results remain consistent with these removals.
[5] We visualize the longitudinal trends of taste exploration based on different time windows in **Extended Data Fig. 7**. Note that our current approach in measuring taste exploration uses prior six-month streams as a source for a user's past taste vector.

**Analysis of Taste Adaptation (Fig. 3a-d; Extended Data Table 2; Supplementary Table 4)**

To capture listeners' taste adaptation to a city, we first generate a user-home city vector, which captures their listening habits at home. Then we calculate the cosine similarity between that user-home vector and the taste vector of the city the user is visiting. To capture city taste distance, we compute the cosine distance between the vector of a user's home city and the vector of a city that the user visited in a given month. The resulting variable, *taste distance between home city and visited city* is our independent variable for these analyses.

**Dependent Variable**

*Taste adaptation to city (monthly).* We measure how much a focal user's taste adapts to the taste of a city she is visiting. Taste adaptation is high if a user's taste assimilates to the visited city's prevalent taste. By contrast, adaptation is low (below zero) if a user's taste moves further from that city's taste when visited. Nevertheless, some users might already possess tastes similar to a city they are visiting. In these instances, even though a user does not adjust her taste to that city, her city-specific vector will appear very similar to the visited city, and taste adaptation will incorrectly appear high. To prevent this, we take into account the baseline similarity between a user and a city in our computation of taste adaptation. With this approach, taste adaptation can vary between -1 and 1. The former (i.e., taste adaptation close to -1) corresponds to a consumer whose taste was similar to a focal city's taste before visiting, but becomes distant from that city while visiting. By contrast, the latter (i.e., taste adaptation close to 1) reflects a consumer whose taste was very different from a focal city before visiting but becomes more similar to that city while there. More formally, user *i*'s taste adaptation to a city *c* at month *t*, $Taste\ Adaptation_{i,c,t}$, is measured as follows:

$$Taste\ Adaptation\ to\ a\ city_{i,c,t} = \cos sim\ (UV_{i,c,t}, CV_c) - \cos sim\ (UV_{i,h}, CV_c)$$

where $UV_{i,c,t}$ is the user vector *i* based on music that user played while she was in a city *c* at month *t*, $CV_c$ is the city taste vector *c*, and $UV_{i,h}$ is the user's taste vector in home city *h*.

**Control Variable**

***Geographical distance to city.*** As our first analysis of taste exploration suggests, it is possible that a user's taste adaptation is also influenced by geospatial change from home to a destination city. We measure the physical distance from a user's home city to a city she visits as the haversine distance between the two.

We also include the same set of covariates as in our analysis of taste exploration. They include ***distance from global taste***, ***song recency***, ***listening count***, ***algorithmic listening***, and ***month dummies***.

**Empirical Strategy and Results**

We again implement a fixed-effects linear model using the *xtreg* framework in Stata 17 to estimate the association between taste distance and a user's taste adaptation. We use robust standard errors clustered on the user, and we standardize all variables before running the model. Descriptive statistics for all variables are reported in **Supplementary Table 1**.

In **Fig. 3a,** we visualize the relationship between city taste distance and user taste adaptation, with insets depicting country-level coefficients. Across the models, we see a positive association between the two, suggesting that the more a visited city's taste environment differs from a user's home city environment, the more that user will adjust their listening preferences toward the place visited (see **Extended Data Table 2**). A similar pattern appears across different countries (see **Supplementary Table 4**). Furthermore, the interaction between taste distance and geospatial distance indicates that the positive relationship between taste distance and adaptation is positively moderated by long-distance travel (Model 8-11 in **Extended Data Table 2** and **Fig. 3c**). These interaction effects are less prominent in the five countries from our sample with fewer users (Model 5-9 in **Supplementary Table 4**).

**Analysis of Lockdown Shock: Differences-in-Differences Estimation (Fig. 4; Extended Data Table 3)**

For our DiD estimation, we examine differences in taste exploration between Australian and South African users before and after the imposition of a nationwide lockdown in South Africa. To test whether a nationwide lockdown changed the taste exploration trajectory of listeners, we estimate:

$$Y_{it} = \beta_0 + \beta_1 treated_i + \beta_2 post_t + \beta_3 treated_i \times post_t + \beta_4 controls_{it-1} + u_i + v_t + \varepsilon_{it}$$

where $Y_{it}$ is the *taste exploration* of listener $i$ in week $t$, $treated_{it}$ equals one if a listener $i$ is in South Africa and zero if he/she is in Australia, and $post_t$ equals one if the current week follows South Africa's first nationwide lockdown date (zero for the week before it).[6] We add

---

[6] Note that, unlike in the previous analyses in which variables are measured on a monthly basis, here we use weekly time windows to more precisely exploit the timing of the nationwide lockdown imposed in South Africa.

*controls*$_{it-1}$, which include five covariates—*algorithmic listening, listening count, song recency, distance from national taste,* and *travel distance*—measured on a weekly basis and lagged one week. We also include both user fixed-effects ($u_i$) and week fixed-effects ($v_t$). $\varepsilon_{it}$ is the error term. We cluster standard errors at the individual listener level. The coefficient of interest is $\beta_3$, which measures the difference in taste exploration ($Y_{it}$) between treated users and control users. While this setting is not the most ideal to evaluate treatment effects and we are hesitant to draw strict causal inference, the result may be interpreted as the suggestive average impact of routine disruption on taste exploration. Descriptive statistics for the variables are reported in **Supplementary Table 1**.

Our model contains 23,124 observations for 1,241 users—619 South African users and 622 Australian users—who physically stayed in their home country for most of the first half of 2020: the 22-week period from the second week of January to the last week of May. South African users are the treated group and Australians the control group. Note that unlike in the previous analyses where variables are measured on a monthly basis, here we use weekly time windows to more precisely exploit the timing of the nationwide lockdown in South Africa. Results show that the average treatment effect on the treated group (ATET) is positive and significant (*p*-value < 0.001), indicating that South African users, who underwent drastic routine disruption due to the national lockdown, consumed more novel music after the lockdown than did Australian users whose routine disruption was less severe. For robustness checks, we also ran several models without the two fixed-effects and/or other covariates in **Extended Table 3**. None of these different model specifications (dropping user and week fixed-effects in Model 6, removing the five covariates in Model 3) changes the result.

To validate the result of our DiD estimate, we examine whether the trajectories of *taste exploration* are parallel for the control and treatment groups prior to the implementation of the treatment. We check what is known as the parallel-trends or common-trends assumption, an important assumption of the DiD model. The parallel-trends assumption for our estimate is supported by a visual diagnostic and a statistical test. First, a graphical diagnostic using "estat trendplots" in the *xtdidregress* postestimation commands in STATA 17 shows the means of the outcome variable over time for both groups, as well as the results of the linear-trends model (**Fig. 4b**). The graph appears to indicate that the parallel-trends assumption is satisfied: prior to the nationwide lockdown in South Africa, the average taste exploration for users in Australia and South Africa followed parallel paths. We also perform a test for the linear pre-treatment trends using "estat ptrends" following *xtdidregress*. The linear-trends model estimates a coefficient for the differences in linear trends prior to treatment, testing the null hypothesis that linear trends are parallel. If that coefficient is 0, the linear pre-treatment trends are parallel. The result indicates that we do not have evidence to reject the null hypothesis of parallel trends ($F$=0.08, $p$-value=0.781). Hence, both the graphical analysis and the statistical test support the parallel-trends assumption.

In addition, we perform a Granger-type causality test to evaluate whether treatment effects are observed prior to the treatment (i.e., nationwide lockdown), using "estat granger" in the postestimation commands of *xtdidregress* in STATA 17. The null hypothesis here is that the coefficients prior to the treatment are jointly 0, meaning that there are no anticipatory effects (i.e., the effects start prior to treatment). The result indicates that we do not have evidence to reject the null of anticipation of treatment ($F$=0.86, $p$-value=0.583), suggesting that the treatment effects in our DiD model are not observed ahead of March 26, 2020.

# EXTENDED DATA

## Extended Figures

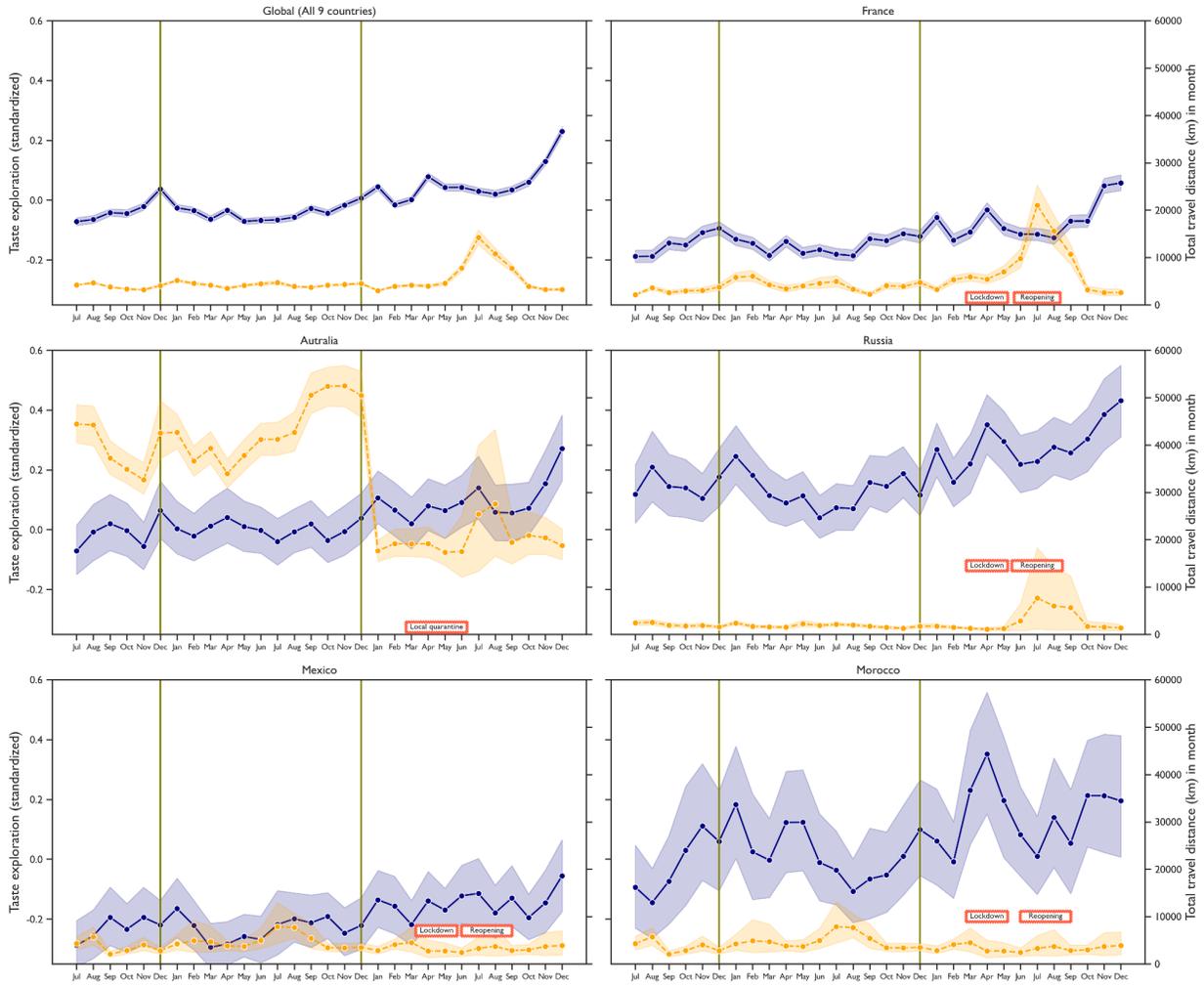

**Extended Data Fig. 1. Taste exploration and travel distance over time.** Taste distances, measured by the cosine of user vectors between the current month and the prior six months. Travel distances measured by the log of haversine-calculated km traveled within the month.

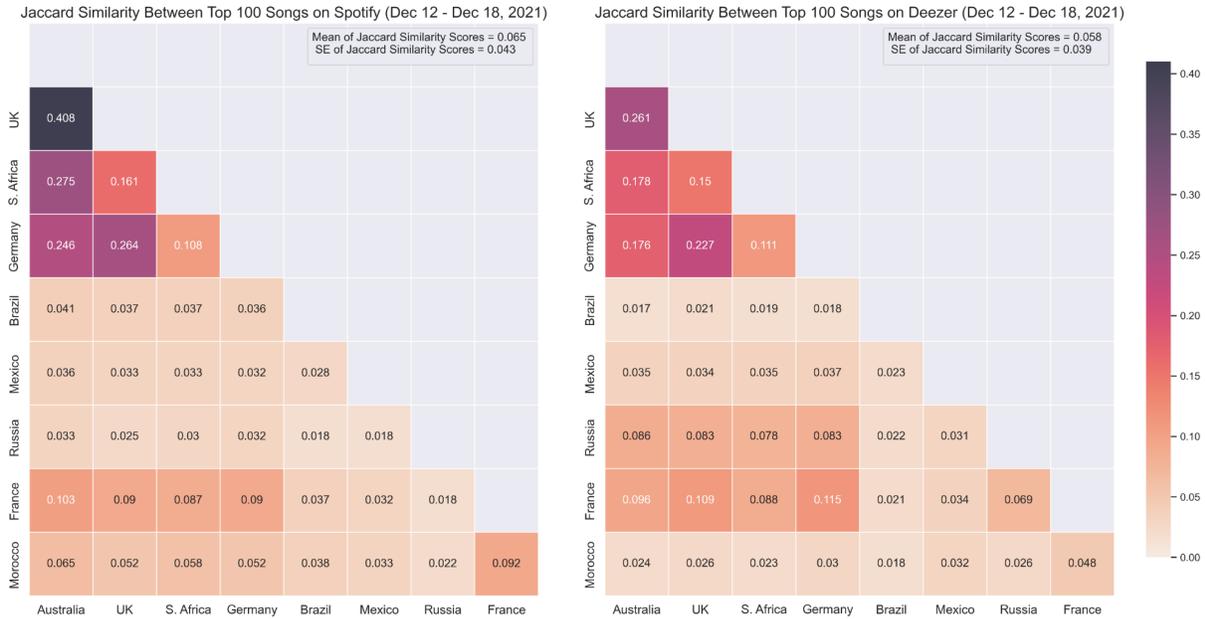

**Extended Data Fig. 2. Jaccard Similarity Between Top 100 Songs on Spotify vs. Deezer.** Cross-country similarity with respect to what is "popular" in each country. Both Spotify and Deezer release daily charts of the most popular songs by country based on number of streams. We aggregated the top 100 songs from Spotify and Deezer during the same 7-day period and measured inter-country similarity by calculating the Jaccard similarity coefficient.

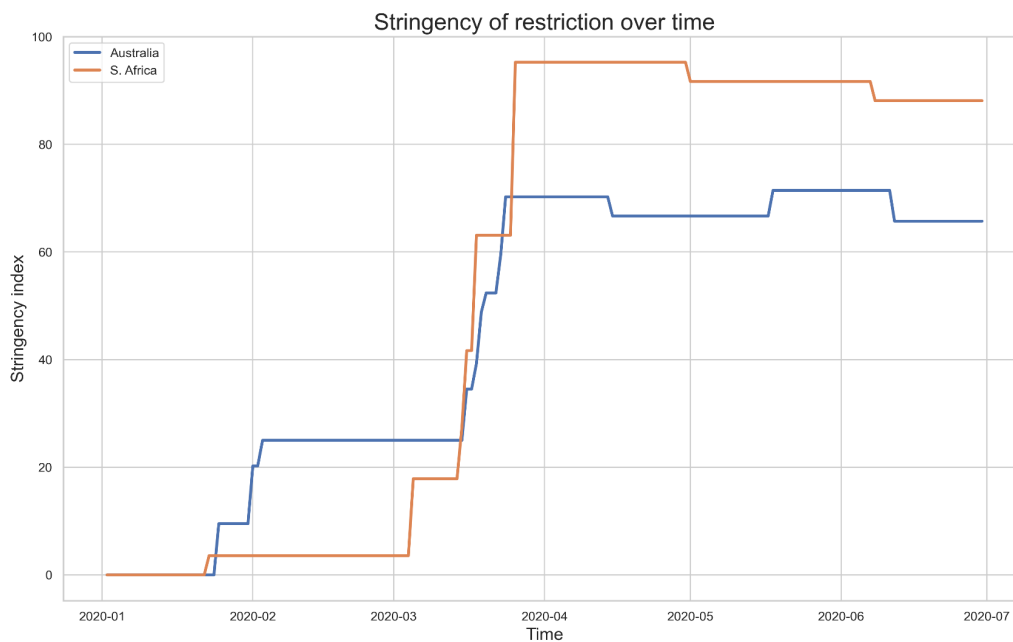

**Extended Data Fig. 3. Stringency of restriction over the first half year of 2020 in South Africa and Australia.** The Oxford Covid-19 Government Response Tracker (OxCGRT) collected systematic information on policy measures that governments implemented to tackle the spread of COVID-19. It tracked various policy responses across different countries from January 1, 2020 to the end of 2022 and quantified governments' policy responses. This includes the stringency index that records the strictness of "lockdown style" policies that primarily restrict people's behavior ranging from 0 (no restriction) to 100 (maximum restriction). The above figure shows daily scores of the stringency index for South Africa and Australia the first half year of 2020. It highlights a drastic increase in stringency in South Africa compared to Australia in late March 2020, but an earlier (and smaller) stringency bump in Australia when it began to close international borders just before February, likely resulting from its proximity to China.

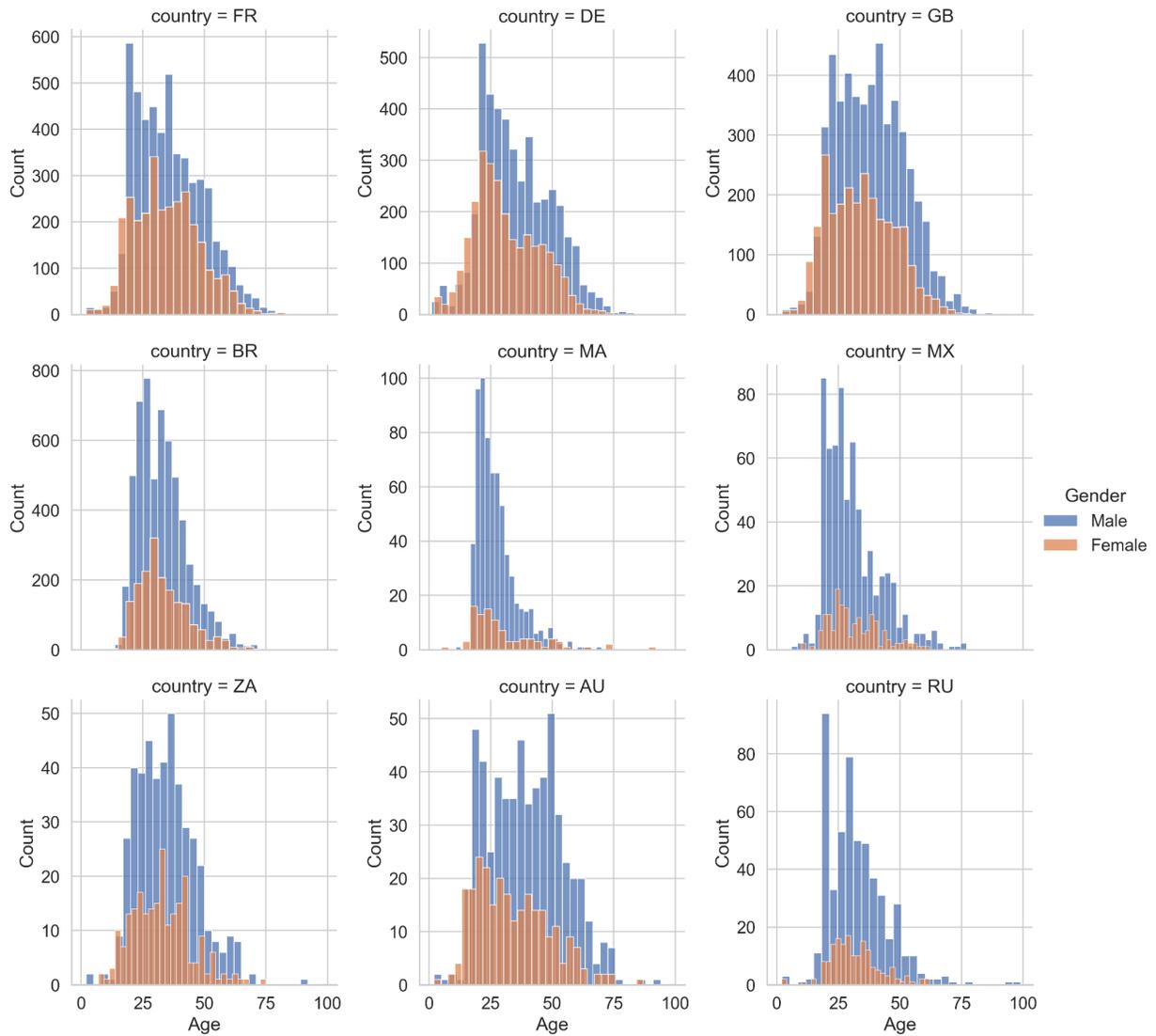

**Extended Data Fig. 4 Distribution of listeners by gender and age.** Distribution of the demographic features of our sampled users in each country. Across all 9 countries, male users outnumber female users, and those aged 20-40 account for the largest portion of the sample.

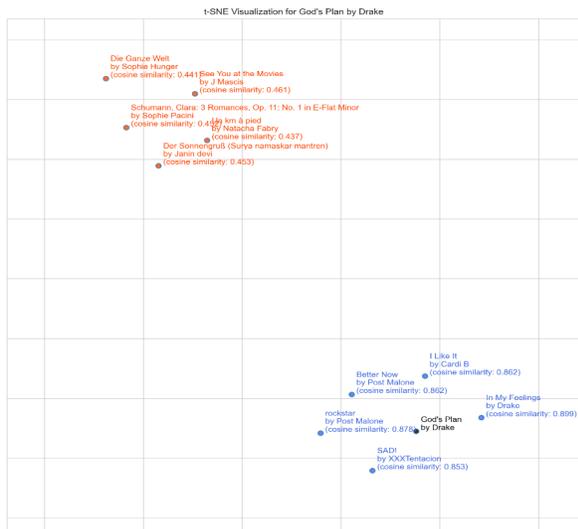
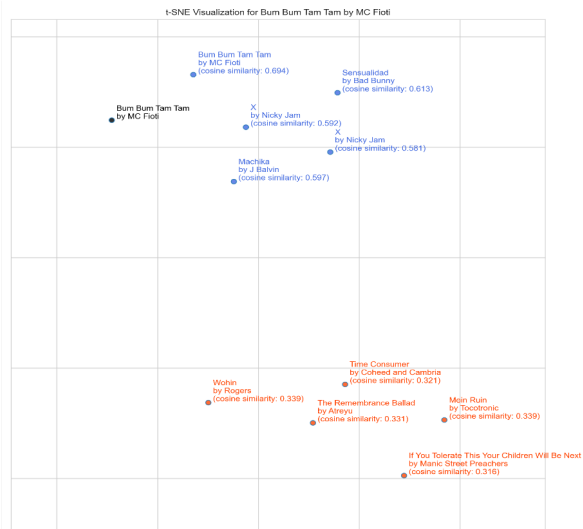
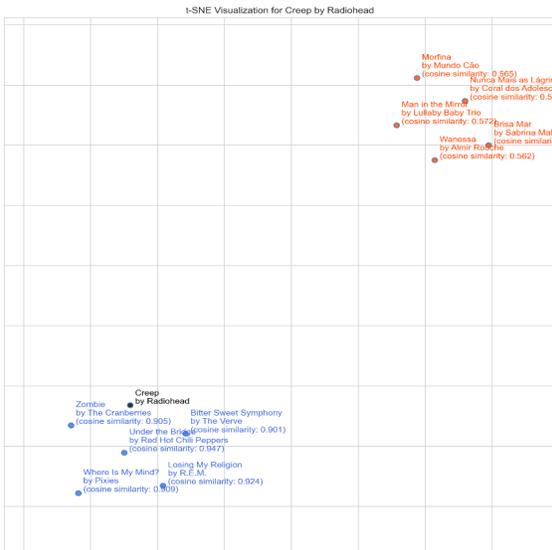
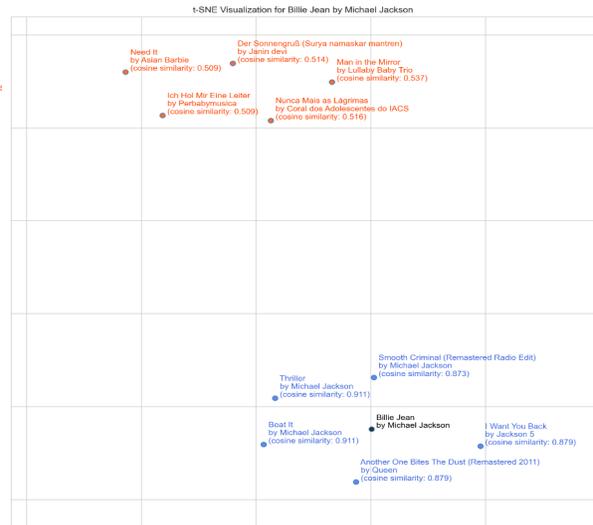
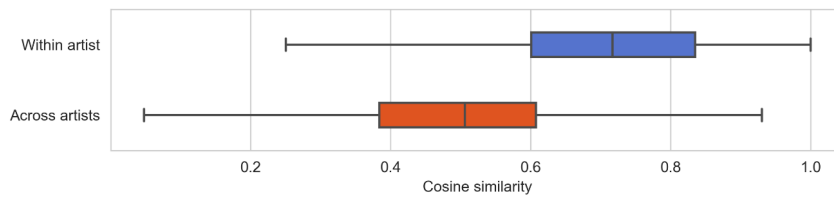

**Extended Data Fig. 5a-e Validation of Song2Vec (S2V).** The 5 most similar and dissimilar songs of a sampled focal song predicted by our S2V model with cosine similarity to that focal song in parentheses. A focal song (in black) in **5a** is *God's Plan* by Drake that debuted at number one on the US Billboard *Hot 100* in January 2018. Its most similar songs are closely located (in blue) and mostly works by other contemporaneous popular musicians in Pop and Hip-Hop. The most dissimilar songs (in red) are distant from Drake's artistic terrain, and include an Indie Folk song by a swiss singer (*Die Ganze Welt* by Sophie Hunger), and one of the *Piano Romance* Op.

11 that were written in 1839 by a female German pianist, Clara Schumann. Concretely, the average cosine similarity of the five most similar songs to *God's Plan* is 0.870 whereas that of the five most dissimilar songs is 0.448. **5b.** Similarly, our model identifies five hit songs by alternative rock bands formed in the 1980-90s as the most similar songs to Radiohead's *Creep*. **5c.** A Brazilian rapper MC FIoti's 2017 song, *Bum Bum Tam Tam*, is positioned closely with other Latin American musicians' hit songs. Note that some of the neighboring songs whose titles are the same as the focal song are different editions of the focal song released in different albums. **5d.** A mega hit by Michael Jackson, *Billie Jean*, has the highest similarity with other famous songs by Michael Jackson or the Jackson 5 of which he was lead member. It is also collocated with Queen's *Another One Bites The Dust*, the British rock band's unusual disco number. **5e.** Lastly, using our S2V model, we compare the average cosine similarity of songs within artists and that of songs across artists. Specifically, the former is the mean of cosine similarities between songs produced by a focal artist (e.g., similarity between all songs by Michael Jackson). The latter is the cosine similarity between a group of songs by an artist and all songs in the population by all the other artists (e.g., similarity between all Michael Jackson songs and the entire collection of other songs by all the other musicians). The average within-artist similarity is 0.715, while the average cross-artist similarity is 0.491.

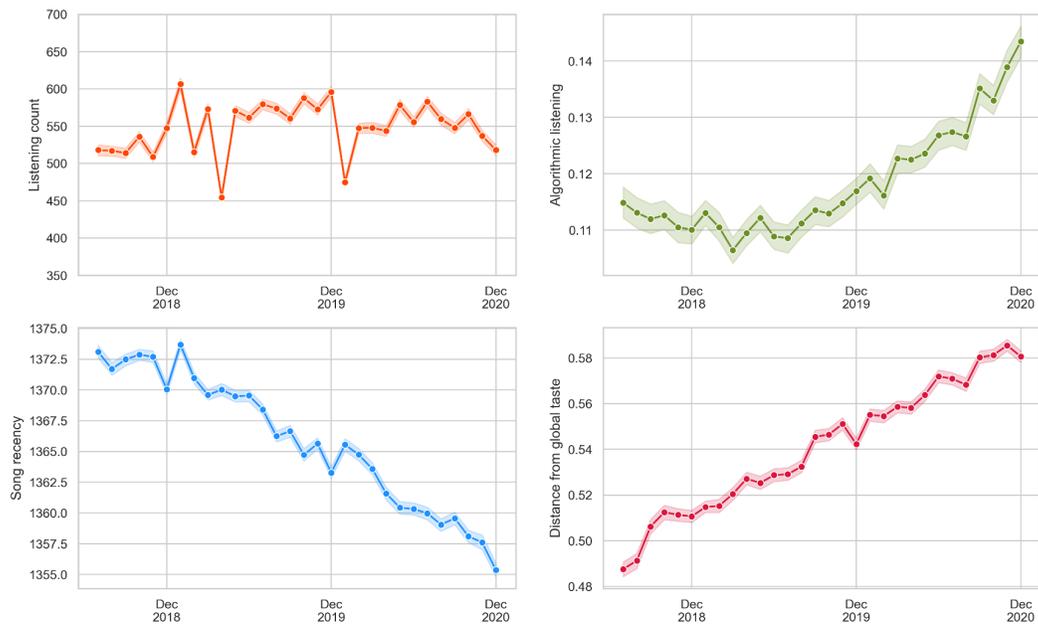

**Extended Data Fig. 6 Longitudinal trend of control variables.** Temporal trend of the mean of each of the four control variables included in our analyses across all users in our data. Listening count—defined as the number of total streams longer than 30 seconds each month—hovers between 450 and 600. Algorithmic listening—ratio of algorithm-driven streams to the total streams by user—takes off in 2020 although it stays below 15%. Song recency—inverse song age—continues to decline while the most considerable drops occur in two Decembers, presumably driven by classic holiday music. Distance from global taste gradually increases overtime although its upward trend substantially drops in two Decembers, which again suggests a holiday effect.

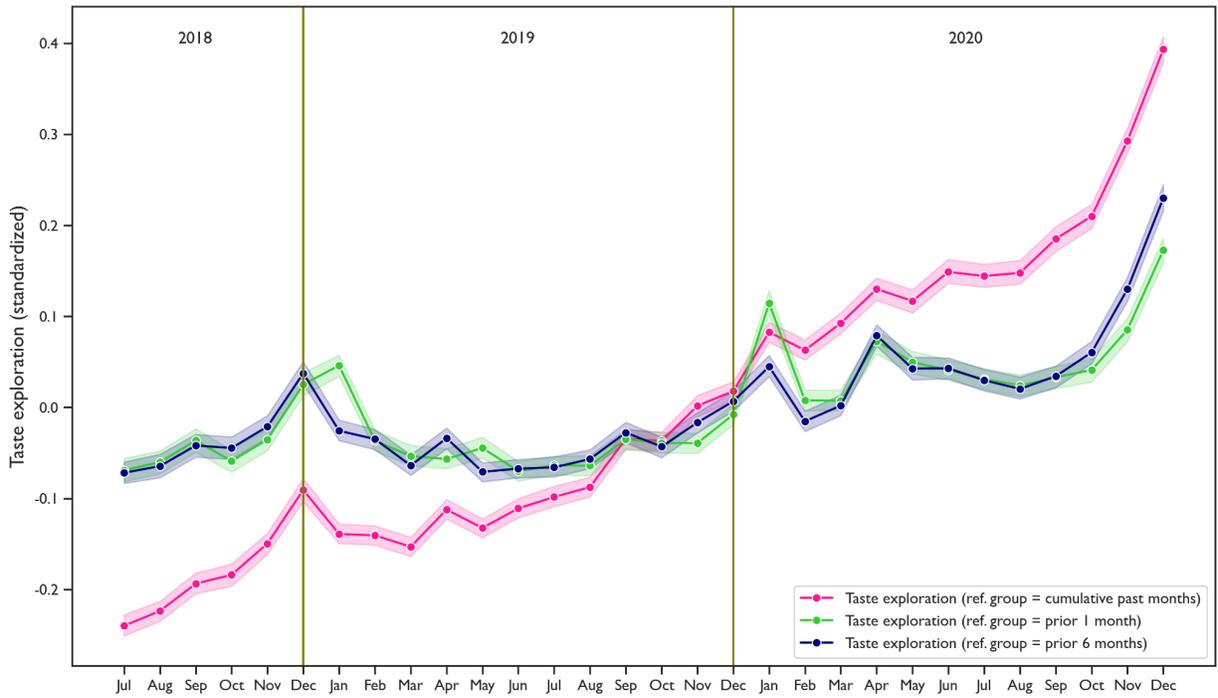

**Extended Data Fig. 7 Longitudinal trend of taste exploration based on different time windows for calculating baseline taste.** We compare taste exploration calculated with respect to the listener's prior month (green), their prior six months (blue), and their entire history of within-platform listening (pink).

# Extended Tables

| | Model 1 (No Cov.) | Model 2 (Months) | Model 3 (Linear) | Model 4 (Quadratic) | Model 5 (No AU) | Model 6 (No BR) | Model 7 (No AU&BR) | Model 8 (Interact) | Model 9 (No AU) | Model 10 (No BR) | Model 11 (No AU&BR) |
|---|---|---|---|---|---|---|---|---|---|---|---|
| | | | | DV = Taste exploration in month $t$ | | | | | | | |
| Constant | -.002*** | -.084*** | -.051*** | -.051*** | -.051*** | .029*** | .030*** | -.051*** | -.052*** | .028*** | .028*** |
| | (.000) | (.005) | (.005) | (.005) | (.006) | (.006) | (.006) | (.005) | (.006) | (.006) | (.006) |
| Algorithmic listening | | | -.061*** | -.061*** | -.061*** | -.063*** | -.063*** | -.061*** | -.061*** | -.063*** | -.063*** |
| | | | (.002) | (.002) | (.002) | (.002) | (.002) | (.002) | (.002) | (.002) | (.002) |
| Listening count | | | -.250*** | -.251*** | -.251*** | -.246*** | -.245*** | -.250*** | -.250*** | -.245*** | -.244*** |
| | | | (.003) | (.003) | (.003) | (.004) | (.004) | (.003) | (.003) | (.004) | (.004) |
| Song recency | | | .087*** | .086*** | .083*** | .105*** | .103*** | .086*** | .084*** | .106*** | .104*** |
| | | | (.004) | (.004) | (.004) | (.004) | (.004) | (.004) | (.004) | (.004) | (.004) |
| Distance from global taste | | | .397*** | .397*** | .396*** | .426*** | .426*** | .397*** | .397*** | .426*** | .426*** |
| | | | (.005) | (.005) | (.005) | (.006) | (.006) | (.005) | (.005) | (.006) | (.006) |
| Travel distance | .046*** | .046*** | .049*** | .072*** | .078*** | .075*** | .083*** | .041*** | .040*** | .042*** | .042*** |
| | (.003) | (.003) | (.003) | (.005) | (.006) | (.006) | (.006) | (.002) | (.002) | (.003) | (.003) |
| Travel distance$^2$ | | | | -.001*** | -.001*** | -.001*** | -.001*** | | | | |
| | | | | (.000) | (.000) | (.000) | (.000) | | | | |
| Travel distance × Distance from global taste | | | | | | | | .020*** | .021*** | .020*** | .022*** |
| | | | | | | | | (.003) | (.003) | (.003) | (.004) |
| N of obs. | 820043 | 820043 | 818612 | 818612 | 801996 | 627033 | 610417 | 818612 | 801996 | 627033 | 610417 |
| N of users | 30384 | 30384 | 30339 | 30339 | 29725 | 22922 | 22308 | 30339 | 29725 | 22922 | 22308 |
| $R^2 between$ | .002 | .002 | .194 | .195 | .194 | .234 | .233 | .194 | .193 | .234 | .233 |
| $R^2 overall$ | .002 | .007 | .147 | .147 | .147 | .166 | .166 | .147 | .147 | .166 | .166 |
| $R^2 within$ | .003 | .010 | .126 | .126 | .126 | .131 | .131 | .127 | .126 | .131 | .131 |
| User FEs | Yes | Yes | Yes | Yes | Yes | Yes | Yes | Yes | Yes | Yes | Yes |
| Month FEs | No | Yes | Yes | Yes | Yes | Yes | Yes | Yes | Yes | Yes | Yes |

Robust standard errors in parentheses are adjusted for clusters in users.
$P$-values correspond to two-tailed tests.
Month dummies are omitted due to space limitation.
* $p < 0.05$, ** $p < 0.01$, *** $p < 0.001$

**Extended Data Table 1. Results of regression analysis on taste exploration at the global level.** Estimates are from fixed-effects OLS regressions. Cluster-robust standard errors are shown in parentheses; *p*-values correspond to two-tailed tests. Month dummies are omitted from the table due to space limitation. Model 1 includes only the main independent variable—travel distance—and user fixed-effects. Model 2 adds month fixed-effects. Model 3 adds control variables. Models 4-7 add the quadratic term of travel distance. The full sample is used in Model 4, Australians are dropped in Model 5; Brazilians are dropped in Model 6; and both Australians and Brazilians dropped in Model 7 (Brazilians have outlier tastes, and Austrians have outlier travel distances). Instead of the quadratic term for travel distance, Models 8-11 include the interaction term between travel distance and "distance" from global taste.

|  | Model 1 (No Cov.) | Model 2 (Months) | Model 3 (Full) | Model 4 (Full) | Model 5 (No AU) | Model 6 (No BR) | Model 7 (No AU&BR) | Model 8 (Full) | Model 9 (No AU) | Model 10 (No BR) | Model 11 (No AU&BR) |
|---|---|---|---|---|---|---|---|---|---|---|---|
|  |  |  |  | DV = Taste adaptation to city |  |  |  |  |  |  |  |
| Constant | -.015*** | .096*** | .019*** | -.070*** | -.066*** | -.121*** | -.117*** | -.006* | -.007* | -.062*** | -.066*** |
|  | (.000) | (.003) | (.003) | (.004) | (.005) | (.005) | (.005) | (.003) | (.003) | (.003) | (.004) |
| Algorithmic listening |  |  | .010*** | .010*** | .010*** | .008*** | .008*** | .010*** | .010*** | .008*** | .008*** |
|  |  |  | (.001) | (.001) | (.001) | (.001) | (.001) | (.001) | (.001) | (.001) | (.001) |
| Listening count |  |  | .028*** | .026*** | .026*** | .024*** | .024*** | .027*** | .027*** | .025*** | .025*** |
|  |  |  | (.002) | (.002) | (.002) | (.003) | (.003) | (.002) | (.002) | (.003) | (.003) |
| Song recency |  |  | .030*** | .025*** | .026*** | .017*** | .018*** | .026*** | .027*** | .019*** | .020*** |
|  |  |  | (.002) | (.002) | (.002) | (.002) | (.002) | (.002) | (.002) | (.002) | (.002) |
| Distance from global taste |  |  | -.211*** | -.212*** | -.210*** | -.229*** | -.227*** | -.212*** | -.209*** | -.229*** | -.226*** |
|  |  |  | (.002) | (.002) | (.002) | (.002) | (.002) | (.002) | (.002) | (.002) | (.002) |
| Geographical distance to city |  |  | .002 | -.000 | .005 | .004 | .014*** | -.020*** | -.030*** | -.020*** | -.034*** |
|  |  |  | (.002) | (.002) | (.003) | (.002) | (.004) | (.002) | (.003) | (.002) | (.004) |
| Taste distance to city | .705*** | .714*** | .717*** | .561*** | .558*** | .563*** | .559*** | .691*** | .691*** | .695*** | .695*** |
|  | (.006) | (.006) | (.006) | (.005) | (.006) | (.006) | (.006) | (.005) | (.005) | (.005) | (.005) |
| Taste distance to city$^2$ |  |  |  | .078*** | .079*** | .078*** | .079*** |  |  |  |  |
|  |  |  |  | (.004) | (.004) | (.004) | (.004) |  |  |  |  |
| Taste distance to city × Geographical distance to city |  |  |  |  |  |  |  | .044*** | .052*** | .050*** | .063*** |
|  |  |  |  |  |  |  |  | (.003) | (.004) | (.004) | (.005) |
| N of obs. | 3429442 | 3429442 | 3429442 | 3429442 | 3333325 | 2873835 | 2777718 | 3429442 | 3333325 | 2873835 | 2777718 |
| N of users | 30254 | 30254 | 30254 | 30254 | 29640 | 22883 | 22269 | 30254 | 29640 | 22883 | 22269 |
| $R^2$ between | .493 | .490 | .664 | .673 | .673 | .723 | .722 | .665 | .663 | .720 | .717 |
| $R^2$ overall | .433 | .438 | .520 | .526 | .528 | .544 | .546 | .521 | .523 | .541 | .543 |
| $R^2$ within | .337 | .346 | .365 | .373 | .374 | .386 | .387 | .368 | .369 | .380 | .382 |
| User FEs | Yes | Yes | Yes | Yes | Yes | Yes | Yes | Yes | Yes | Yes | Yes |
| Month FEs | No | Yes | Yes | Yes | Yes | Yes | Yes | Yes | Yes | Yes | Yes |

Robust standard errors in parentheses are adjusted for clusters in users.
$P$-values correspond to two-tailed tests.
Month dummies are omitted due to space limitation.
* $p < 0.05$, ** $p < 0.01$, *** $p < 0.001$

**Extended Data Table 2. Results of regression analysis on taste adaptation at the global level.** Estimates are from fixed-effects OLS regressions. Cluster-robust standard errors are shown in parentheses; *p*-values correspond to two-tailed tests. Month dummies are omitted from the table due to space limitation. Model 1 includes only the main independent variable—taste distance to city—and user fixed-effects. Model 2 adds month fixed-effects. Model 3 adds control variables. Models 4-7 add the quadratic term of taste distance to city; the full sample used in Model 4, Australians dropped in Model 5; Brazilians dropped in Model 6; both Australians and Brazilians dropped in Model 7 (see ED Table 1). Instead of the quadratic term for taste distance to city, Models 8-11 include the interaction term between taste distance to city and geographical distance to city.

|  | DV = Taste exploration at $t$ week | | | | | | |
| --- | --- | --- | --- | --- | --- | --- | --- |
|  | Model 1 | Model 2 | Model 3 | Model 4 | Model 5 | Model 6 | Model 7 |
| Constant | -.4800*** (.0160) | -.5837*** (.0130) | -.4921*** (.0166) | -.5766*** (.0051) | -.5267*** (.0073) | -.5210*** (.0074) | -.5766*** (.0051) |
| Algorithmic listening |  |  |  | -.0063** (.0024) | -.0060* (.0024) | -.0062** (.0024) | -.0064** (.0024) |
| Listening count |  |  |  | .0018 (.0023) | .0021 (.0022) | .0020 (.0022) | .0017 (.0023) |
| Song recency |  |  |  | -.1063*** (.0041) | -.1107*** (.0040) | -.1107*** (.0039) | -.1062*** (.0041) |
| Distance from nat'l taste |  |  |  | .3429*** (.0046) | .3479*** (.0043) | .3478*** (.0043) | .3427*** (.0046) |
| Travel distance |  |  |  | .0028* (.0012) | .0027* (.0012) | .0026* (.0012) | .0028* (.0012) |
| TREATED (S. Africa) | -.1674*** (.0244) |  | -.1821*** (.0255) |  | -.1068*** (.0112) | -.1176*** (.0117) |  |
| POST (Mar 26 - May 31, 2020) |  | .0416*** (.0053) | .0250*** (.0072) |  | .0114*** (.0031) | -.0003 (.0043) | -.0046 (.0075) |
| TREATED × POST |  |  | .0315** (.0105) |  |  | .0223*** (.0061) | .0225*** (.0061) |
| N of obs. | 23124 | 23124 | 23124 | 23123 | 23123 | 23123 | 23123 |
| N of users | 1241 | 1241 | 1241 | 1241 | 1241 | 1241 | 1241 |
| $R^2$ between | .037 | .002 | .038 | .801 | .813 | .813 | .798 |
| $R^2$ overall | .026 | .002 | .028 | .761 | .772 | .772 | .758 |
| $R^2$ within | .000 | .007 | .008 | .653 | .652 | .653 | .653 |
| User FEs | No | No | No | Yes | No | No | Yes |
| Week FEs | No | No | No | Yes | No | No | Yes |

Standard errors in parentheses are adjusted for 1,241 clusters in users.

$P$-values correspond to two-tailed tests.

The number of users in the control group (i.e., Australians in Australia) is 622.

The number of users in the treated group (i.e., S. Africans in S. Africa) is 619.

Time range of the observations is from the 2nd week of January to the last week of May in 2020.

Week dummies are omitted due to space limitation.

* $p < 0.05$, ** $p < 0.01$, *** $p < 0.001$

**Extended Data Table 3. Results of Difference-in-Differences (DiD) of taste exploration between South Africa and Australia.** Estimates are from fixed-effects OLS regressions. Cluster-robust standard errors are shown in parentheses; $p$-values correspond to two-tailed tests; out of 1,241 users, 622 are Australians in Australia, and 619 are South Africans in South Africa. Week dummies are omitted from the table due to space limitations. Models 1 and 2 include only the treatment dummy and only the dummy for post-treatment periods, respectively. Model 3 shows the average treatment effect on the treated (ATET) without considering control variables and fixed-effects. Models 4-7 include control variables. The result of Model 6 intimates a

positive ATET with control variables and without fixed-effects. Model 7 shows the positive ATET when considering both control variables and fixed-effects.

# SUPPLEMENTARY INFORMATION

**Descriptive statistics for variables used in regression of taste *exploration***

| Variable | Obs | Mean | Std. Dev. | Min | Max |
|---|---|---|---|---|---|
| Taste exploration | 824,621 | 0 | 1 | -.871 | 9.725 |
| Algorithmic listening | 841,010 | 0 | 1 | -.824 | 1.78 |
| Listening count | 841,010 | 0 | 1 | -4.478 | 3.124 |
| Song recency | 848,281 | 0 | 1 | -32.051 | 2.819 |
| Dist. from glob. taste | 826,089 | 0 | 1 | -2.111 | 3.784 |
| Travel distance | 841,010 | 0 | 1 | -.116 | 83.054 |

All variables are measured at the user-month level and are standardized before running regression. Algorithmic listening and listening count are log-transformed before standardized.

**Descriptive statistics for variables used in regression of taste *adaptation***

| Variable | Obs | Mean | Std. Dev. | Min | Max |
|---|---|---|---|---|---|
| Taste adaptation | 3,429,442 | -.021 | .989 | -3.115 | 4.468 |
| Taste distance | 3,429,442 | -.008 | .979 | -.646 | 5.031 |
| Geospatial distance | 3,429,442 | -.028 | .97 | -.451 | 5.717 |
| Algorithmic listening | 3,429,442 | -.071 | .887 | -.486 | 4.453 |
| Listening count | 3,429,442 | .377 | 1.18 | -.862 | 19.57 |
| Song recency | 3,429,442 | .175 | .911 | -12.22 | 2.819 |
| Dist. from glob. taste | 3,429,442 | -.036 | .984 | -2.111 | 3.784 |

All variables are measured at the user-city level and are standardized before running regression. Means do not converge to zero as some datapoints are omitted because of cities without taste vector (e.g., when a user visits small cities where we have no sampled users living there and thus no city-vector available).

**Descriptive Statistics for variables used in diff-in-diffs (South Africa, Treated)**

| Variable | Obs | Mean | Std. Dev. | Min | Max |
|---|---|---|---|---|---|
| Taste exploration | 12598 | -.653 | .524 | -3.06 | .305 |
| Travel distance | 12598 | -.132 | .268 | -.187 | 12.995 |
| Algorithmic listening | 12598 | -.122 | .838 | -.476 | 3.387 |
| Listening count | 12598 | -.049 | .825 | -.818 | 10.301 |
| Song recency | 12598 | .171 | .939 | -6.939 | 1.695 |
| Dist. from glob. taste | 12598 | -.015 | .95 | -1.509 | 3.046 |
| TREATED | 12598 | 1 | 0 | 1 | 1 |
| POST | 12598 | .489 | .5 | 0 | 1 |

All variables are measured at the user-week level and are standardized before running regression.

**Descriptive Statistics for variables used in diff-in-diffs (Australia, Control)**

| Variable | Obs | Mean | Std. Dev. | Min | Max |
|---|---|---|---|---|---|
| Taste exploration | 11602 | -.494 | .457 | -1.906 | .337 |
| Travel distance | 11602 | .121 | .677 | -.187 | 14.361 |
| Algorithmic listening | 11602 | .079 | 1.061 | -.476 | 3.387 |
| Listening count | 11602 | .12 | 1.118 | -.818 | 10.525 |
| Song recency | 11601 | -.186 | 1.031 | -7.308 | 1.69 |
| Dist. from glob. taste | 11602 | .016 | 1.051 | -1.564 | 3.07 |
| TREATED | 11602 | 0 | 0 | 0 | 0 |
| POST | 11602 | .49 | .5 | 0 | 1 |

All variables are measured at the user-week level and are standardized before running regression.

**Supplementary Table 1: Summary statistics for variables in regression analyses**

|  | DV = Taste exploration in month $t$ | | | | | | | | |
| --- | --- | --- | --- | --- | --- | --- | --- | --- | --- |
|  | Model 1 (FR) | Model 2 (UK) | Model 3 (DE) | Model 4 (BR) | Model 5 (RU) | Model 6 (MA) | Model 7 (AU) | Model 8 (MX) | Model 9 (ZA) |
| Constant | -.056*** | .057*** | .140*** | -.277*** | .040 | .096 | .075 | -.204*** | -.011 |
|  | (.010) | (.011) | (.014) | (.010) | (.043) | (.075) | (.043) | (.036) | (.032) |
| Algorithmic listening | -.062*** | -.067*** | -.065*** | -.052*** | -.075*** | -.073*** | -.062*** | -.017 | -.049*** |
|  | (.003) | (.003) | (.004) | (.003) | (.013) | (.021) | (.014) | (.013) | (.009) |
| Listening count | -.236*** | -.222*** | -.262*** | -.270*** | -.307*** | -.291*** | -.263*** | -.313*** | -.234*** |
|  | (.006) | (.007) | (.008) | (.005) | (.022) | (.043) | (.022) | (.030) | (.017) |
| Song recency | .100*** | .160*** | .068*** | .024*** | .116*** | .118 | .201*** | .173*** | .126*** |
|  | (.007) | (.008) | (.008) | (.007) | (.031) | (.065) | (.033) | (.034) | (.024) |
| Distance from global taste | .416*** | .503*** | .398*** | .278*** | .300*** | .492*** | .433*** | .266*** | .488*** |
|  | (.010) | (.011) | (.011) | (.010) | (.030) | (.069) | (.037) | (.043) | (.031) |
| Travel distance | .069*** | .120*** | .067*** | .065*** | .087** | .303 | .019 | .093 | .131* |
|  | (.007) | (.013) | (.017) | (.012) | (.027) | (.214) | (.014) | (.087) | (.051) |
| Tabel distance$^2$ | -.001*** | -.002** | -.001 | -.000 | -.000 | -.023 | .000 | -.000 | -.002* |
|  | (.000) | (.001) | (.001) | (.000) | (.000) | (.031) | (.000) | (.007) | (.001) |
| N of obs. | 183608 | 169364 | 210198 | 191579 | 15534 | 4039 | 16616 | 8105 | 19569 |
| N of users | 6682 | 6219 | 7621 | 7417 | 605 | 155 | 614 | 304 | 722 |
| $R^2$ between | .237 | .239 | .198 | .139 | .172 | .155 | .302 | .197 | .148 |
| $R^2$ overall | .178 | .190 | .131 | .119 | .140 | .165 | .195 | .146 | .145 |
| $R^2$ within | .161 | .179 | .096 | .120 | .125 | .201 | .147 | .141 | .186 |
| User FEs | Yes | Yes | Yes | Yes | Yes | Yes | Yes | Yes | Yes |
| Month FEs | Yes | Yes | Yes | Yes | Yes | Yes | Yes | Yes | Yes |

Robust standard errors in parentheses are adjusted for clusters in users.

$P$-values correspond to two-tailed tests.

Month dummies are omitted due to space limitation.

\* $p < 0.05$, \*\* $p < 0.01$, \*\*\* $p < 0.001$

**Supplementary Table 2. Results of regression analysis on taste exploration at the country level (curvilinear relationship).** Estimates from fixed-effects OLS regressions. Cluster-robust standard errors are shown in parentheses; $p$-values correspond to two-tailed tests. Month dummies are omitted from the table due to space limitations. Each model corresponds to a separate regression analysis on each of the nine countries. Model 1 (FR) uses the same model specification but only with samples from French users. Model 2 (UK) also uses the same model specification but with British users, and so on. A most salient pattern that emerges across the models is the positive coefficient of travel distance, although the coefficients for the three countries with small sample size are relatively small and statistically not significant. In addition, coefficients for squared travel distance are significantly negative only for France, United the Kingdom, and South Africa. This leads us to argue that it is more appropriate to see the relationship between travel distance and taste exploration as linear with a concave, diminishing effect rather than as an inverted-U.

|  | DV = Taste exploration in month $t$ | | | | | | | | |
|---|---|---|---|---|---|---|---|---|---|
|  | Model 1 (FR) | Model 2 (UK) | Model 3 (DE) | Model 4 (BR) | Model 5 (RU) | Model 6 (MA) | Model 7 (AU) | Model 8 (MX) | Model 9 (ZA) |
| Constant | -.052*** | .060*** | .147*** | -.285*** | .030 | .094 | .072 | -.204*** | -.020 |
|  | (.010) | (.011) | (.014) | (.011) | (.044) | (.075) | (.042) | (.037) | (.033) |
| Algorithmic listening | -.064*** | -.069*** | -.065*** | -.053*** | -.076*** | -.074*** | -.063*** | -.017 | -.050*** |
|  | (.003) | (.003) | (.004) | (.003) | (.014) | (.022) | (.014) | (.013) | (.009) |
| Listening count | -.230*** | -.212*** | -.259*** | -.265*** | -.303*** | -.277*** | -.254*** | -.311*** | -.231*** |
|  | (.006) | (.007) | (.008) | (.005) | (.022) | (.048) | (.022) | (.030) | (.016) |
| Song recency | .106*** | .170*** | .070*** | .031*** | .121*** | .126 | .205*** | .179*** | .130*** |
|  | (.007) | (.008) | (.008) | (.007) | (.031) | (.071) | (.033) | (.037) | (.024) |
| Similarity to global taste | .425*** | .507*** | .407*** | .277*** | .302*** | .501*** | .427*** | .266*** | .484*** |
|  | (.010) | (.011) | (.011) | (.010) | (.030) | (.077) | (.038) | (.043) | (.031) |
| Travel distance | .056*** | .021*** | .052*** | .026*** | .009 | .036 | .007 | .003 | .023*** |
|  | (.006) | (.004) | (.009) | (.004) | (.013) | (.050) | (.004) | (.009) | (.006) |
| Travel distance × Distance from global taste | .047*** | .008 | .035*** | .032*** | -.007 | .038 | .003 | .002 | .019** |
|  | (.007) | (.004) | (.010) | (.006) | (.017) | (.054) | (.004) | (.010) | (.007) |
| N of obs. | 183608 | 169364 | 210198 | 191579 | 15534 | 4039 | 16616 | 8105 | 19569 |
| N of users | 6682 | 6219 | 7621 | 7417 | 605 | 155 | 614 | 304 | 722 |
| $R^2$ between | .228 | .233 | .197 | .130 | .169 | .151 | .299 | .200 | .148 |
| $R^2$ overall | .173 | .185 | .131 | .113 | .136 | .161 | .192 | .146 | .144 |
| $R^2$ within | .159 | .175 | .096 | .116 | .121 | .196 | .145 | .140 | .185 |
| User FEs | Yes | Yes | Yes | Yes | Yes | Yes | Yes | Yes | Yes |
| Month FEs | Yes | Yes | Yes | Yes | Yes | Yes | Yes | Yes | Yes |

Robust standard errors in parentheses are adjusted for clusters in users.
$P$-values correspond to two-tailed tests.
Month dummies are omitted due to space limitation.
* $p < 0.05$, ** $p < 0.01$, *** $p < 0.001$

**Supplementary Table 3. Results of regression analysis on taste exploration at the country level (interaction effects).** Estimates are from fixed-effects OLS regressions. Cluster-robust standard errors are shown in parentheses; $p$-values correspond to two-tailed tests. Month dummies are omitted from the table from space limitations. Travel distance squared in **Supplementary Table 2** is replaced by the interaction term between travel distance and distance from global taste. Although not across all countries, three out of four large-sample countries show highly significant coefficients of the interaction term, in addition to South Africa.

|  | | | | DV = Taste adaptation to city | | | | | |
|---|---|---|---|---|---|---|---|---|---|
|  | Model 1 (FR) | Model 2 (UK) | Model 3 (DE) | Model 4 (BR) | Model 5 (RU) | Model 6 (MA) | Model 7 (AU) | Model 8 (MX) | Model 9 (ZA) |
| Constant | -.008 | -.060*** | -.154*** | .252*** | .044 | -.172*** | -.170*** | .177*** | -.065* |
|  | (.009) | (.005) | (.007) | (.007) | (.039) | (.045) | (.015) | (.026) | (.028) |
| Algorithmic listening | .011*** | .008*** | .003 | .020*** | .005 | -.004 | .016*** | .026** | .011 |
|  | (.002) | (.002) | (.002) | (.002) | (.011) | (.010) | (.005) | (.009) | (.008) |
| Listening count | .014* | .036*** | .026*** | .042*** | .060*** | .023* | .023*** | .046*** | .036*** |
|  | (.006) | (.004) | (.002) | (.005) | (.016) | (.010) | (.006) | (.007) | (.009) |
| Song recency | .001 | -.025*** | .051*** | .060*** | -.022 | .096*** | -.029** | -.017 | .021 |
|  | (.007) | (.004) | (.002) | (.005) | (.016) | (.027) | (.009) | (.027) | (.012) |
| Distance from global taste | -.208*** | -.294*** | -.203*** | -.101*** | -.169*** | -.259*** | -.298*** | -.136*** | -.326*** |
|  | (.006) | (.005) | (.003) | (.005) | (.019) | (.021) | (.008) | (.016) | (.013) |
| Geographical distance to city | -.113*** | -.010** | -.030* | -.023*** | -.166*** | -.072* | -.016*** | -.038* | -.043*** |
|  | (.012) | (.003) | (.014) | (.004) | (.033) | (.035) | (.002) | (.017) | (.007) |
| Taste distance to city | .776*** | .668*** | .657*** | .647*** | .708*** | .708*** | .709*** | .609*** | .727*** |
|  | (.008) | (.009) | (.007) | (.014) | (.029) | (.028) | (.025) | (.029) | (.036) |
| Taste distance to city × Geographical distance to city | .077*** | .039*** | .086*** | .022*** | .050 | .024 | -.003 | .012 | .004 |
|  | (.009) | (.007) | (.010) | (.005) | (.025) | (.024) | (.006) | (.018) | (.016) |
| N of obs. | 848664 | 686372 | 1137084 | 555607 | 23254 | 25407 | 96117 | 20579 | 36358 |
| N of users | 6676 | 6209 | 7614 | 7371 | 591 | 155 | 614 | 304 | 720 |
| $R^2$ between | .756 | .798 | .684 | .726 | .540 | .765 | .794 | .612 | .785 |
| $R^2$ overall | .589 | .562 | .488 | .515 | .528 | .466 | .500 | .418 | .562 |
| $R^2$ within | .458 | .357 | .334 | .281 | .408 | .312 | .359 | .315 | .410 |
| User FEs | Yes | Yes | Yes | Yes | Yes | Yes | Yes | Yes | Yes |
| Month FEs | Yes | Yes | Yes | Yes | Yes | Yes | Yes | Yes | Yes |

Robust standard errors in parentheses are adjusted for clusters in users.

$P$-values correspond to two-tailed tests.

Month dummies are omitted due to space limitation.

* $p < 0.05$, ** $p < 0.01$, *** $p < 0.001$

**Supplementary Table 4. Results of regression analysis on taste adaptation at the country level.** Estimates are from fixed-effects OLS regressions. Cluster-robust standard errors are shown in parentheses; *p*-values correspond to two-tailed tests. Month dummies are omitted from the table due to space limitation. The positive coefficient of taste distance to city appears highly significant across all models. Its interaction with geospatial distance to city is also significantly positive for the four big-sample countries—France, United Kingdom, Germany, and Brazil.